\documentclass[superscriptaddress,amssymb,amsmath,nobibnotes,aps,prd,eqsecnum,showpacs,
nofootinbib]{revtex4}
\usepackage{titlesec}
\usepackage{mathtools}
\usepackage{graphicx}

\usepackage{multirow}
\usepackage[usenames,dvipsnames,svgnames]{xcolor}

\usepackage[english]{babel}

\usepackage{amsmath}

\usepackage{amssymb}

\usepackage{amsfonts}

\usepackage{subfig}

\usepackage{dcolumn}
\usepackage{epsf}
\usepackage{enumerate}
\usepackage{hhline}
\usepackage{array}
\usepackage{tabularx}
\usepackage{appendix}
\usepackage[colorlinks=true,
            linkcolor=blue,
          urlcolor=gray,
            citecolor=red]{hyperref}
\usepackage{epstopdf}
\newcommand{\be}{\begin{equation}}

\newcommand{\ee}{\end{equation}}

\newcommand{\bea}{\begin{eqnarray}}

\newcommand{\eea}{\end{eqnarray}}

\newcommand{\beaa}{\begin{eqnarray*}}

\newcommand{\eeaa}{\end{eqnarray*}}

\newcommand{\nn}{\nonumber \\}

\newcommand {\pd}{\partial}

\begin{document}

\title{Inflation in mimetic $f(G)$ gravity}

  \author{Yi Zhong}
 \affiliation{Institut de Ci\`{e}ncies de l'Espai (ICE, CSIC), Carrer de Can Magrans s/n, Campus UAB, 08193 Bellaterra (Barcelona), Spain}
\affiliation{Institut d'Estudis Espacials de Catalunya (IEEC), 08034 Barcelona, Spain}
\affiliation{Institute of Theoretical Physics,
            Lanzhou University, \\Lanzhou 730000,
            People's Republic of China}
\author{Diego S\'aez-Chill\'on G\'omez}
\affiliation{Institut de Ci\`{e}ncies de l'Espai (ICE, CSIC), Carrer de Can Magrans s/n, Campus UAB, 08193 Bellaterra (Barcelona), Spain}
\affiliation{Institut d'Estudis Espacials de Catalunya (IEEC), 08034 Barcelona, Spain}


\begin{abstract}
Mimetic gravity is analysed in the framework of some extensions of General Relativity, where a function of the Gauss-Bonnet invariant in four dimensions is considered. By assuming the so-called mimetic condition, the conformal degree of freedom is isolated and a pressureless fluid naturally arises. Then, the complete set of field equations for mimetic Gauss-Bonnet gravity is established and some inflationary models are analysed, for which the corresponding gravitational action is reconstructed. The spectral index and tensor-to-scalar ratio are obtained and compared with observational bounds from Planck and BICEP2/Keck array data. The full agreement with above data is achieved for several versions of the mimetic Gauss-Bonnet gravity. Finally, some extensions of Gauss-Bonnet mimetic gravity are considered and the possibility of reproducing inflation is also explored.
\end{abstract}

\maketitle

\section{Introduction}
Since the eighties inflation has been widely studied as a certain epoch occurred during a very early stage of the universe evolution, when a very short but super-accelerating phase transformed a microscopic universe into a macroscopic one, solving some theoretical problems of the Big Bang model concerning the initial conditions of the universe. In addition, inflation does not produce a perfectly symmetric universe, since quantum fluctuations grow rapidly because of the effects of the rapid expansion, becoming macroscopic perturbations. This is one of the main success of the inflationary scenario, since these quantum fluctuations form the primordial seeds for all the large structure created at later times in the universe as well as the anisotropies observed in the Cosmic Microwave Background (CMB), for a review see \cite{Mukhanov:1990me,Mukhanov}. Some alternatives have been raised since then, as the ekpyrotic scenario, but all of them including a super-accelerating phase as inflation (see \cite{Khoury:2001wf} and references therein).  \\

In addition, over the last decade observations from missions as Wilkinson Microwave anisotropy probe (WMAP) \cite{Peiris:2003ff} and the recent Planck mission and BICEP2 \cite{Planck-Inflation}, have provided a way to measure the spectral index of the power spectrum of primordial perturbations produced during inflation and the ratio of the tensor and scalar perturbations, drawing again much attention on the differences among the many existing inflationary models. Particularly, inflation is well described by the so-called slow-roll models, usually described by a single scalar field that mimics a cosmological constant during the inflationary epoch, rolling down the step of its potential by the end of inflation, when the field is assumed to decay in different particles that reheat the universe, recovering the initial state of Big Bang theory \cite{Mukhanov}. Actually the form of the potential for the scalar field can be related to the spectral index of the power spectrum for the scalar perturbations generated during inflation as well as to the the tensor perturbations, such that the appropriate form of the scalar potential can be well reconstructed departing from the observational data \cite{Lidsey:1995np}. \\

Alternatively to the usual scalar field models for inflation, some extensions of General Relativity (GR) have been widely analysed not only to stage as an alternative to the dark energy problem but also as a realistic candidate for driving inflation  (for a review see \cite{Reviews}). In particular, the so-called $f(R)$ gravities, an extension of the Hilbert-Einstein action, have been considered as a serious candidate for the inflationary epoch. Particularly, some of the most promising inflationary models are constructed within the $f(R)$ gravity scenario, since some of these models can easily reproduce slow-roll inflation by mimicking an effective cosmological constant during the inflationary epoch and then decaying, leading to the desirable behaviour and the right values for spectral index of the power spectrum for scalar perturbations and the tensor to scalar ratio \cite{Bamba:2014wda}. Actually one of the most popular inflationary model within modifications of GR is the so-called $R^2$ inflation or Starobinsky model \cite{Starobinsky:1980te}, which assumes a quadratic extra term in the the Hilbert-Einstein action, leading to a nearly scale invariant power spectrum and a negligible tensor-to-scalar ratio, as predicted by the last data released from Planck \cite{Planck-Inflation}. In addition, any deviation from Starobinsky inflation should be small in order to avoid deviations from its well known results, as suggested by some recent analysis \cite{delaCruz-Dombriz:2016bjj}. In fact, $f(R)$ gravities can also be extended to provide a successful description for the whole cosmological evolution (see Refs~\cite{Cognola:2007zu}), in particular the so-called exponential gravity \cite{Odintsov:2017qif}. Some other extensions of GR also include other invariants in the action, as powers of the Riemann and Ricci tensors or non-standard couplings to the matter Lagrangian and the energy-momentum tensor \cite{Harko:2011kv}. A special mention deserves the so-called Gauss-Bonnet gravities, where a non-linear function $f(\cal{G})$ of the Gauss-Bonnet invariant is included in the action \cite{Gauss-Bonnet}. Some cosmological solutions have been explored, including the exact $\Lambda$CDM model \cite{Elizalde:2010jx} and some inflationary models \cite{Kanti:2015pda,Oikonomou:2015qha}, since an accelerating expansion can be easily achieved. Nevertheless, the presence of non-linear Gauss-Bonnet terms in the action may introduce ghost instabilities in an empty anisotropic universe, i.e. the Kasner-type background, although such instabilities are absent on Friedmann-Lema\^itre-Robertson-Walker (FLRW) cosmologies (see \cite{DeFelice}).\\


On the other hand, several observational proofs suggest that besides the possible dark energy content, the majority matter in our universe is composed by an unknown fluid that seems to interact just gravitationally (or at least  interact very weakly with standard matter) and behaves as a pressureless fluid in the universe expansion, the so-called cold dark matter. Though there are many descriptions of dark matter, the nature of the dark matter is still unknown.  Some models describe dark matter as a particle (for a review see \cite{Feng:2010gw}), while some attempts are focused on modifications of newtonian dynamics at certain scales \cite{Sanders:2002pf}. Nevertheless, other promising mechanics for dark matter are encompassed under the name of mimetic gravity, which provides a geometric description for dark matter. The original version of mimetic gravity is obtained by isolating the conformal degree of freedom of GR by an auxiliary scalar field, leading to the same behaviour as a pressureless fluid after integrating the equations \cite{Chamseddine:2013kea,Chamseddine:2014vna}. Other equivalent formulations introduce a constraint directly in the gravitational action through a Lagrange multiplier \cite{Golovnev:2013jxa} or analyses the lack of invariance under ``disformations'' for the mimetic case \cite{Deruelle:2014zza}. Some extensions of original mimetic gravity have been discussed, where the conformal degree of freedom is isolated also in theories as $f(R)$ or $f(\cal{G})$ gravities, which provides a complete solution to the dark matter and dark energy problems, as shown in Refs.~\cite{Nojiri:2014zqa,Momeni:2014qta,Astashenok:2015haa,Nojiri:2016vhu,Leon:2014yua,Odintsov:2015wwp}. As commented above, some modified gravities have shown to provide a consistent explanation for dark energy and inflation, such that the mimetic version of such theories may give a correct description of dark matter as well (for a review on modified mimetic gravities see \cite{Sebastiani:2016ras}). Particularly, inflation and the dynamics at the early universe has been analysed in some extensions of mimetic gravity \cite{Nojiri:2016vhu,Myrzakulov:2015qaa,Bouhmadi-Lopez:2017lbx} as well as the growth of cosmological perturbations \cite{Matsumoto:2015wja}. Moreover, in Ref.~\cite{Astashenok:2015haa}, the mimetic version for $f({\cal G})$ gravity was investigated and bouncing cosmology can be realised. In addition, other classical aspects of mimetic gravities have been explored, as the existence of spherically symmetric solutions, as black holes \cite{Deruelle:2014zza,Chen:2017ify}. \\

 In this paper we will investigate the possibility of reproducing inflation in mimetic $f({\cal G})$ gravity. Several inflationary models are studied and the corresponding gravitational action is reconstructed. Then, the viability of such models is analysed by confronting their predictions to the last data by Planck and BICEP2/Keck Array. In addition, some extensions of mimetic gravity are also studied, where the auxiliary scalar field becomes dynamical by adding a kinetic term and a potential in its Lagrangian. Different approaches are assumed to reconstruct the appropriate inflationary solutions and their respective gravitational actions. To that aim, we assume several ansatzs for the scale factor and then, the corresponding gravitational Lagrangian is reconstructed. Such method has been widely used in the literature as an alternative to find exact solutions in higher order theories of gravity \cite{Capozziello2002}, an always difficult task due to the complexity of the field equations. Then, generalisations of General Relativity have been reconstructed which contain some of the most important solutions in cosmology, as for instance exact $\Lambda$CDM in $f(R)$ gravity \cite{Alvaro} or in Gauss-Bonnet extensions \cite{Elizalde:2010jx}.\\

This paper is organised as follows: in section \ref{mimetic} we introduce mimetic $f({\cal G})$ gravity. Section \ref{inflation} is devoted to the analysis of inflation in the original mimetic gravity with a function of the Gauss-Bonnet invariant. After that, section \ref{extensions} deals with some extensions of mimetic gravity where the scalar field acquires a dynamical behaviour. Finally, the conclusions of the paper are summarised in section \ref{conclusions}.

\section{Mimetic f(G) gravity}
\label{mimetic}
Mimetic gravity is constructed to isolate the conformal degree of freedom by expressing the physical metric in terms of an auxiliary metric and an scalar field as follows \cite{Chamseddine:2013kea},
 \begin{eqnarray}
        g_{\mu\nu}=-\hat{g}^{\rho\sigma} \pd_{\rho}\phi\pd_{\sigma}\phi \hat{g}_{\mu\nu}\ .
        \label{mimetic_gauge}
    \end{eqnarray}
It is straightforward to show that the physical metric turns out invariant under the conformal transformation $\hat{g}_{\mu\nu}\rightarrow \Omega^2\hat{g}_{\mu\nu}$, while from (\ref{mimetic_gauge}), the following constraint on the scalar field is obtained:
\be
g^{\mu\nu}\partial_{\mu}\phi\partial_{\nu}\phi=-1\ .
\label{Constraint}
\ee
Then, in mimetic gravity, Hilbert-Einstein action is assumed in terms of the physical metric $g_{\mu\nu}$, 
\be
S=\frac{1}{2\kappa^2}\int dx^4 \sqrt{-g}R(g_{\mu\nu})\ ,
\label{EH_action}
\ee
While the field equations can be expressed by varying the action with respect to metric and writing such variation in terms of the variation of the auxiliary metric and the  the scalar field (\ref{mimetic_gauge}), leading to:
\bea
(G^{\mu\nu}-T^{\mu\nu})+(G-T)g^{\mu\lambda}g^{\mu\gamma}\partial_{\lambda}\phi\partial_{\gamma}\phi&=&0\ , \nn
\nabla_{\mu}\left[(G-T)\partial^{\mu}\phi\right]&=&0\ .
\label{mimetic_eqs}
\eea
Here the field equations are expressed solely in terms of the physical metric $g_{\mu\nu}$ and the scalar field that accounts for the conformal degree of freedom. The key point here arises because the mimetic field behaves as an effective pressureless fluid, such that can be interpreted as a contribution to dark matter \cite{Chamseddine:2013kea,Chamseddine:2014vna}. The appearance of a new dynamical degree of freedom arises because the variation to make the action stationary assumes less conditions \cite{Golovnev:2013jxa}. Nevertheless, the action (\ref{EH_action}) is not unique and can be extended to provide also an explanation to dark energy, as done by considering $f(R)$ gravity and isolating again the conformal degree of freedom \cite{Nojiri:2014zqa}. \\  
 
In this manuscript, we are interested to extend mimetic gravity by considering a non-linear function of the Gauss-Bonnet topological invariant through different approaches and to study the viability of reproducing inflation in this framework and study its predictions. Hence, let us start by recalling the general gravitational action for the so-called $f(G)$ gravity, 
    \begin{eqnarray}
        S=\int d^4x\sqrt{-g}\left[ \frac{R}{2\kappa^2} + f(\mathcal{G}) \right] + S_m,
        \label{action_mgb1}
    \end{eqnarray}
where
    \begin{eqnarray}
        \mathcal{G}=R^2-4R_{\mu\nu}R^{\mu\nu}+R_{\mu\nu\lambda\sigma}R^{\mu\nu\lambda\sigma}
        \label{action gb term}
    \end{eqnarray}
is the Gauss-Bonnet term, a topological invariant in 4 dimensions. Here it follows that $R_{\mu\nu}$ and $R_{\mu\nu\lambda\sigma}$ are the Ricci and Riemann tensors respectively. For simplicity, in the following we are assuming natural units $\kappa^2=8\pi G/c^4=1$. Then, by parametrising $g_{\mu\nu}$ in terms the auxiliary scalar field $\phi$ as in (\ref{mimetic_gauge}), the field equations can be obtained by using the auxiliary metric $\hat{g}_{\mu\nu}$ as in the original mimetic case (\ref{EH_action}), leading to the following field equations,
\begin{eqnarray}
&&R_{\mu\nu}- \frac{1}{2}R g_{\mu\nu}+ \left(f_{\cal G}{\cal G}-f({\cal G})\right)g_{\mu\nu}
+   \nonumber \\
&&8\Big[R_{\mu\rho\nu\sigma}+R_{\rho\nu}g_{\sigma\mu}-R_{\rho\sigma}g_{\nu\mu}-R_{\mu\nu}g_{\sigma\rho}+R_{\mu\sigma}g_{\nu\rho}
+\frac{R}{2}\left(g_{\mu\nu}g_{\sigma\rho}-g_{\mu\sigma}g_{\nu\rho}\right)\Big]\nabla^{\rho}\nabla^{\sigma}f_{\cal
G}  \nonumber \\
&&+\partial_{\mu}\phi\partial_{\nu}\phi\left(-R+8\left(-R_{\rho\sigma}
+\frac{1}{2}R
g_{\rho\sigma}\right)\nabla^{\rho}\nabla^{\sigma}f_{\cal
G}+4(f_{\cal G}{\cal G}-f({\cal G}))\right)=T_{\mu\nu}+\partial_{\mu}\phi\partial_{\nu}\phi T,
\label{eom_metric}\\
&&\nabla^{\mu}\left(\partial_{\mu}\phi\left(-R+8\left(-R_{\rho\sigma}+\frac{1}{2}
R g_{\rho\sigma}\right)\nabla^{\rho}\nabla^{\sigma}f_{\cal
G}+4(f_{\cal G}{\cal G}-f({\cal G})\right)-T)\right)=0.
\label{eom_scalar}
\end{eqnarray}
As in the original action for mimetic gravity (\ref{mimetic_eqs}), the field equations do not depend on the auxiliary metric but on the physical one $g_{\mu\nu}$ and on the scalar field $\phi$, which encompasses the conformal degree of freedom as above. Since we are interested in studying spatially-flat Friedmann-Robertson-Walker (FRW) spacetimes, we assume the  following form for the metric:
    \begin{eqnarray}
        ds^2=-dt^2+a(t)^2 \delta _{ij}dx^i dx^j,
        \label{metric}
    \end{eqnarray}
While the the scalar curvature $R$ and the Gauss-Bonnet terms are
    \begin{eqnarray}
        R=6\left(\frac{\ddot{a}}{a}+\frac{\dot{a}^2}{a^2}\right)=6(\dot{H}+2H^2),\quad
        {\cal G}=-24\frac{\ddot{a}\dot{a}^2}{a^3}=-24H^2(\dot{H}+H^2). \label{gbinvariant}
    \end{eqnarray}
Hence, the FLRW equation (\ref{eom_metric}) leads to:
\be
8H^2\frac{\partial^2 f_{\cal G}}{\partial t^2}+16H\left(\dot{H}+H^2\right)\frac{\partial f_{\cal G}}{\partial t}-\left(f_{\cal G}{\cal G}-f\right)+2\dot{H}+3H^2=-p\ .
\label{FLRW1GB}
\ee
Whereas by integrating the equation for the scalar field (\ref{eom_scalar}), we obtain
\be
4H^2\frac{\partial^2 f_{\cal G}}{\partial t^2}+4H\left(2\dot{H}+3H^2\right)\frac{\partial f_{\cal G}}{\partial t}+\frac{2}{3}\left(f_{\cal G}{\cal G}-f\right)+\dot{H}+2H^2=-\frac{C}{a^3}-\frac{\rho}{6}+\frac{p}{2}\ ,
\label{Scalar1GB}
\ee
where $C$ is a integration constant. Here we have assumed a perfect fluid for the energy-momentum tensor $T_{\mu\nu}=(\rho+p)u_{\mu}u_{\nu}+pg_{\mu\nu}$. The first term in the r.h.s of (\ref{Scalar1GB}) arises naturally after integrating the equation (\ref{eom_scalar}), representing the same behaviour as a pressureless perfect fluid, or in other words, representing the so-called mimetic dark matter. Then, by combining both equations (\ref{FLRW1GB}) and (\ref{Scalar1GB}), the following equation is obtained:
    \begin{eqnarray}
        4H^2 \frac{dg(t)}{dt}+4H(2\dot{H}-H^2)g(t)=B(t)\ ,
        \label{eom_gt}
    \end{eqnarray}
where $B(t)=-\dot{H}-\frac{1}{2}(\rho+p)-\frac{C}{a^3}$, and $g(t)$ is defined as $g(t) \equiv \frac{df_{\mathcal{G}}}{dt}$. Hence, for a particular ansatz for the Hubble parameter, the equation (\ref{eom_gt}) can be solved for $g(t)$ and using the expression (\ref{gbinvariant}), the corresponding mimetic Gauss-Bonnet action can be reconstructed. For instance, whether we consider a universe filled with only dust and mimetic dark matter, this implies $p=0$ and $\rho=\rho_0 a(t)^{-3}$, and therefore $B(t)$ is reduced to $B(t)=-\dot{H}-\tilde{C}a(t)^{-3}$ and $\tilde{C}\equiv C+\frac{\rho_0}{2}$, such that the general solution for $g(t)$ is:
    \begin{eqnarray}
        g(t)=\left[\frac{H_0}{H(t)}\right]^2 \text{exp}\left(\int_{0}^{t}H(t_1)dt_1\right)
            \left[ g_0+\frac{1}{4H_0^2} \int_{0}^{t} \text{exp}\left(-\int_{0}^{t_2}H(t_1)dt_1\right) B(t_2)dt_2 \right]\ .
        \label{gt}
    \end{eqnarray}
And for the appropriate Hubble parameter, the gravitational action is obtained. In the following, we consider several inflationary scenarios, and the corresponding $f({\cal G})$ is reconstructed.

\section{Inflation in mimetic Gauss-Bonnet gravity}
\label{inflation}

We are interested in studying slow-roll inflation within the framework of mimetic Gauss-Bonnet gravities, expressed by the action (\ref{action_mgb1}). In common slow-roll inflationary models, the responsible of the super-acceleration phase is an scalar field, which can be characterised by the following Lagrangian:
\be
\label{2.1}
S_{\phi} = \int {\rm d}^4 x \sqrt{-g} \left[- \frac{1}{2}\partial_\mu \phi \partial^\mu \phi - V(\phi) \right]\, ,
\ee
The corresponding FLRW equations for such an action are:
\bea
\frac{3}{\kappa^2} H^2 = \frac{1}{2}{\dot \phi}^2 + V(\phi)\, ,  \nn
- \frac{1}{\kappa^2} \left( 3 H^2 + 2\dot H \right)
= \frac{1}{2}{\dot \phi}^2 - V(\phi)\, ,
\label{2.2}
\eea
whereas the scalar field equation is given by:
\be
\ddot{\phi}+3H\dot{\phi}+\frac{\partial V(\phi)}{\partial\phi}=0
\label{2.3}
\ee
For convenience, we use the number of e-folds $N=ln(\frac{a(t)}{a(0)})$ as the independent variable instead of the cosmic time $t$. Moreover, the scalar field can be redefined as $\phi=N$, such that the above equations (\ref{2.2}) can be rewritten as follows:
\bea
\omega(\phi) &=& -\frac{2 H'(\phi)}{\kappa^2 H(\phi)}\, , \nn
V(\phi) &=&  \frac{1}{\kappa^2}\left[ 3 \left(H(\phi)\right)^2 + H(\phi) H'(\phi) \right]\, .
\label{2.6}
\eea
Here $\omega(\phi)$ is the kinetic term for the scalar field that arises when redefining the scalar field. In slow-roll inflationary models with an scalar field, the field behaves effectively as a cosmological constant during inflation, while the Hubble parameter is approximately constant, such that $H\dot{\phi}\gg\ddot{\phi}$ and $V\gg\dot{\phi}^2$. At the end of inflation, the field rolls down decaying in particles, reheating the universe. Nevertheless, during inflation fluctuations of the scalar field produces a fast growth of the curvature and tensor perturbations, whose characteristic amplitudes and scale
dependence are related to the scalar filed Lagrangian through the scalar potential $V(\phi)$. By defining the so-called slow-roll parameters:
\be
\epsilon=
\frac{1}{2\kappa^2} \left( \frac{V'(\phi)}{V(\phi)} \right)^2\, ,\quad
\eta= \frac{1}{\kappa^2} \frac{V''(\phi)}{V(\phi)}\, , \quad
\lambda^2 = \frac{1}{\kappa^4} \frac{V'(\phi) V'''(\phi)}{\left(V(\phi)\right)^2}\, .
\label{2.4}
\ee
The spectral index $n_\mathrm{s}$ of the curvature perturbations and the tensor-to-scalar ratio $r$ can be expressed in terms of the slow-roll parameters (\ref{2.4}) as follows
\be
n_\mathrm{s} - 1= - 6 \epsilon + 2 \eta\, ,\quad r = 16 \epsilon \, , \quad \alpha_\mathrm{s} = \frac{{\rm d} n_\mathrm{s}}{{\rm d}\log k} \sim 16\epsilon \eta - 24 \epsilon^2 - 2 \xi^2\, .
\label{2.5}
\ee
Then, by using the equations (\ref{2.6}), the slow-roll parameters $\epsilon$ and $\eta$ can be expressed in terms of the Hubble parameter \cite{Bamba:2014wda},
    \begin{eqnarray}
        \label{slow_roll_11}
        \epsilon=-\frac{H(N)}{4H'(N)}\left[
        \frac{H''(N)H(N)+6H'(N)H(N)+H'^2(N)}{3H^2(N)+H'(N)H(N)}\right]^2\\
        \label{slow_roll_12}
        \eta=-\frac{\left(9\frac{H'(N)}{H(N)}+3\frac{H''(N)}{H(N)}+\frac{1}{2}\left(
        \frac{H'(N)}{H(N)}\right)^2-\frac{1}{2}\left(
        \frac{H''(N)}{H'(N)}\right)^2+3
        \frac{H''(N)}{H'(N)}+\frac{H'''(N)}{H'(N)}\right)}{2\left(3+\frac{H'(N)}{H(N
        )}\right)}\ ,
    \end{eqnarray}
where the primes denote derivatives with respect to the number of e-folds $N$. Hence, by using the above expressions, we can calculate the corresponding predictions for a particular inflationary model and compare to the recent constraints provided by the Planck collaboration (see Ref.~\cite{Planck-Inflation}):
\begin{equation}\label{constraintedvalues}
n_s=0.968\pm 0.006\, , \quad r<0.07\, .
\end{equation}
Note that here we are assuming that our model (\ref{action_mgb1}) mimics well slow-roll inflation, as the extra terms in the action can be considered as a perfect fluid and the mimetic field just enters in the equations through a pressureless fluid. Other approaches where the curvature perturbation is directly obtained from the Gauss-Bonnet field equations have been previously considered in \cite{Oikonomou:2015qha}. Finally, by using the tools from Section \ref{mimetic}, the corresponding mimetic action is obtained. In the following, we study some examples, where inflation occurs and the gravitational action is reconstructed.

\subsection{Example 1}
Firstly let us study a power-law inflation model, whose cosmological evolution is given by 
\be
a(t)=a_0 t^n\ .
\ee
The Hubble parameter is given by $H(t)= \frac{n}{t}$, which can be equivalently expressed in terms of the number of e-folds as:
    \begin{eqnarray}
       H(N)= n \text{e}^{-\frac{N}{n}}\ .
       \label{inflation1_Hn}
    \end{eqnarray}
Note that the Hubble parameter $H(t)= \frac{n}{t}$ is divergent at $t=0$, when the initial singularity occurs. By using Eq. (\ref{gt}), we obtain
    \begin{eqnarray}
        g(t)=t^{n+1}\left[ g_1 - \frac{-1+t^{-n}+\frac{\tilde{C}(-1+t^{2-4n})}{2-4n}}{4n^2} \right] ,
        \label{gt3}
    \end{eqnarray}
Recalling that $g(t)= f_{\cal G}(t)$, the above expression yields:
    \begin{eqnarray}
        f_{\cal G}(t) &=& \int g(t) dt    \nonumber\\
                   &=& -\frac{t^2}{8n^2}
                   -\frac{\tilde{C}-2(2n-1)(1+4g_1 n^2)}{8n^2(2+n)(2n-1)}t^{n+2}
                   +\frac{\tilde{C}}{8n^2(4-3n)(2n-1)}t^{4-3n}\ .
        \label{gt13}
    \end{eqnarray}
And by using the expression for the Gauss-Bonnet term (\ref{gbinvariant}), we can express $f_{\cal G}$ as a function of the Gauss-Bonnet invariant $\cal G$:
    \begin{eqnarray}
        f_{\cal G}\left({\cal G}\right) = f_1 {\cal G}^{-\frac{1}{2}} + f_2 {\cal G}^{-\frac{n+2}{4}}
        +f_3 {\cal G}^{\frac{3n}{4}-1}\ ,
    \end{eqnarray}
where
    \begin{eqnarray}
        f_1&=& -\frac{1}{2}\sqrt{\frac{3(1-n)}{2n}},\nonumber\\
        f_2&=&
        \frac{2^{\frac{3}{4}(n-2)} 3^{\frac{2+n}{4}}
         [n^3 (1-n)]^{\frac{2+n}{4}}[-\tilde{C}+2(2n-1)(1+4g_1 n^2)]}
        {n^2(2+n)(2n-1)},\nonumber\\
        f_3 &=& \frac{2^{-\frac{9n}{4}} 3^{1-\frac{3n}{4}}\tilde{C}n(n-1)[n^3(1-n)]^{-\frac{3n}{4}}}
        {4-11n+6n^2}.
        \nonumber
    \end{eqnarray}
Finally, the corresponding action $f(\cal G)$ is obtained:
     \begin{eqnarray}
        \label{f(G)}
        f\left(\cal G\right) = 2f_1 {\cal G}^{\frac{1}{2}} + \frac{4}{2-n}f_2 {\cal G}^{\frac{-n+2}{4}}
        +\frac{4}{3n}f_3 {\cal G}^{\frac{3n}{4}}.
    \end{eqnarray}
For this model, the slow-roll parameters given by Eqs. (\ref{slow_roll_11}) and (\ref{slow_roll_12})  read,
    \begin{eqnarray}
        \epsilon=\frac{1}{n},   ~~~~~
        \eta=\frac{2}{n}.
    \end{eqnarray}
Therefore the spectral index of primordial curvature
perturbations $n_s$ and the scalar-to-tensor ratio $r$ are
    \begin{eqnarray}
        n_s\simeq1-\frac{2}{n},   ~~~~~
        r=\frac{16}{n}
    \end{eqnarray}
Then, by using the last constraints provided by Planck and BICEP2 (\ref{constraintedvalues}), the following constraint on the parameter $n$ is obtained:
 \be
n>\frac{16}{0.07}\ ,
 \ee
 which leads to an spectral index given by $n_s>0.9916$, outside from the $1\sigma$ confidence region from the joint analysis of Planck and BICEP2. Keeping just the constraint on the tensor-to-scalar ratio provided by Planck collaboration $r<0.12$, the spectral index is restricted to be $n_s>0.985$, still away from the error bars of the spectral index given in (\ref{constraintedvalues}). Hence, we may conclude that the mimetic gravity described by the action (\ref{action_mgb1}) and Eq. (\ref{f(G)}) can not support a power-law inflation is not in full agreement with the Planck and BICEP2/Keck Array data.

\subsection{Example 2}

\begin{figure}[h!]
\centering
\includegraphics[scale=0.6]{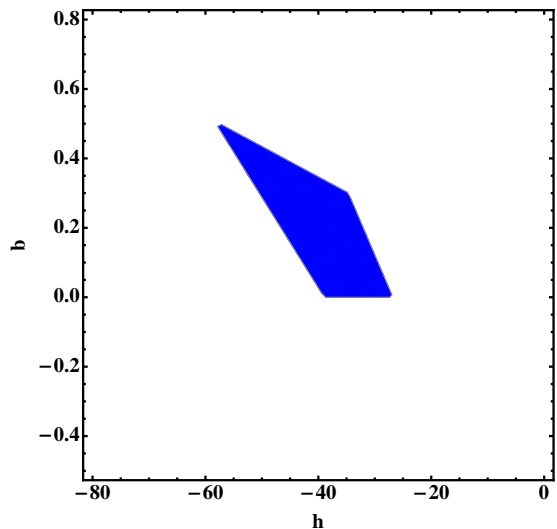}
\caption{Confident region for the parameters values of $h=\frac{G_1}{G_0}+N$ and $b$ to be consistent with the Planck data and BICEP2/Keck Array, $h$ and $b$ are constrained inside the blue region.}
\label{inf2}
\end{figure}

Let us now consider another inflationary model in which the cosmological evolution is described by the Hubble rate,
    \begin{eqnarray}
       H^2 (N)= (G_0 N+G_1)^{b}\ .
       \label{inflation1_Hn}
    \end{eqnarray}
Here $\{G_0, G_1, b\}$ are constants, where the constants $G_0<0$, $G_1>0$. During the inflationary period we have $|\dot{H}|\ll H^2$ which yields $\frac{G_1}{G_0}\gg N$, such that the Hubble parameter is approximately constant (de Sitter) along inflation. After inflation ends, the first term in (\ref{inflation1_Hn}) becomes important and the Hubble rate decays, since $G_0<0$. 
In order to proceed to reconstruct the corresponding mimetic Gauss-Bonnet Lagrangian, we rewrite Eq. (\ref{eom_gt}) in terms of the number of e-folds:
    \begin{eqnarray}\label{transfefold}
        4H(N)^3\partial_N[H(N)g(N)]+4H(N)^3[2H'(N)-H(N)]g(N)=B(N)\ .
    \end{eqnarray}
Here $B(N)=-H(N)H'(N)-\frac{\tilde{C}}{a(N)}$. For simplicity we define $h=\frac{G_1}{G_0}+N$, such that the solution for $g(h)$ is given by:
\begin{eqnarray}\label{transfefold}
       g(h)=g_1(h)\left(g_0+\int \frac{g_2(h)}{g_1(h)}dh\right)\ ,
       \label{gex2}
   \end{eqnarray}
where
    \begin{eqnarray}
       g_1(h)&=&\text{e}^{\int (1-\frac{3b}{h})}\ , \\
       g_2(h)&=&-\frac{(G_0 h)^{-4b}\left[\tilde{C}\text{e}^{h}h+b(G_0 h)^{2b}\right]}{4h}\ .
   \end{eqnarray}
   While the Gauss-Bonnet term leads to:
\be
{\cal G}=-12G_{0}^{2b}h\left(b+2h\right)\ .
\label{GBex2}
\ee   
By integrating equation (\ref{gex2}), the function $g(h)$ yields:
\be
g(h)=\frac{\partial f({\cal G})}{\partial t}=(G_0h)^{b/2}\frac{\partial h}{\partial N}\frac{\partial f({\cal G})}{\partial h}=G_0^{-2b}h^{-3b}\text{e}^{h}\left[g_0G_0^{2b}-\frac{\tilde{C}G_0^{-2b}h^{1-b}}{4(1-b)}+\frac{b}{4}\ \Gamma(b; h)\right]\ .
\label{solex2}
\ee
Here $\Gamma$ is the gamma function. Nevertheless, the explicit form for $f({\cal G})$ can not be obtained exactly, as the above expression (\ref{solex2}) has to be integrated. However, we can analyse the predictions of such model  by combining Eqs. (\ref{slow_roll_11}), (\ref{slow_roll_12}) and (\ref{inflation1_Hn}), the slow-roll parameters read,
    \begin{eqnarray}
        \epsilon&=&-\frac{b(-1+b+6h)^2}{2h(b+6h)^2}\ ,\\
        \eta&=&-\frac{3-b(5-2b-12h)-6h}{2h(b+6h)}\ .
    \end{eqnarray}
Thus the spectrum index $n_s$ and the scalar-to-tensor ratio $r$ become:
    \begin{eqnarray}
        n_s&=&1+\frac{b-1}{h}-\frac{18}{(b+6h)^2}+\frac{12}{b+6h}\ ,\\
        r&=&-\frac{8b(b+6h-1)^2}{h(b+6h)^2}\ .
    \end{eqnarray}
Here the inflationary parameters for the scalar and tensor fluctuations depend on the number of e-folds but also on the value of the free parameters, particularly on $b$ and on the combination $h=G_1/G_0+N$, so in order to obtain some constrains for the values of $b$ and $h$, we use the Planck and BICEP2/Keck Array data as above. The allowed region for the free parameters is shown in Fig.~\ref{inf2}, which shows the $1\sigma$ region for the spectral index, while the values for the tensor-to-scalar ratio satisfy the constraint (\ref{constraintedvalues}) within the blue region of Fig.~\ref{inf2}. In addition, the values for the free parameters within the $1\sigma$ region allow to have a tensor-to-scalar ratio as small as required, for instance when considering $h=-40$ and $b=0.0001$, this leads to $r\sim 10^{-5}$.

\subsection{Example 3}
\begin{figure}
\centering
    \includegraphics[scale=0.6]{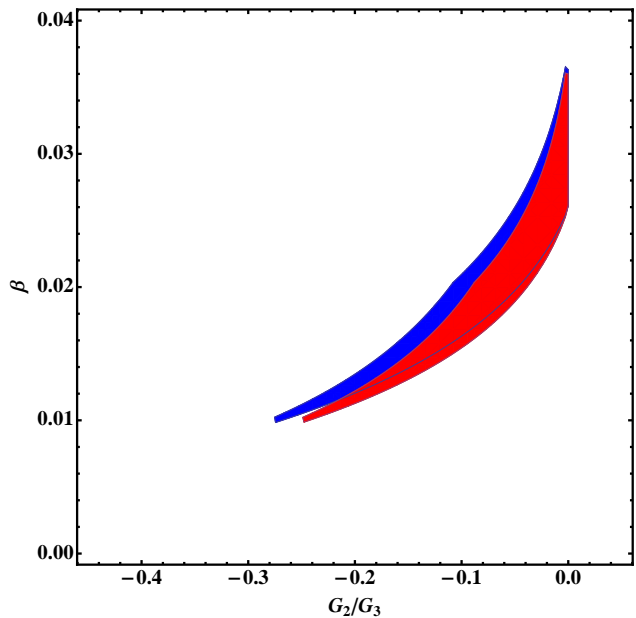}
\caption{Constrains for the parameters $\frac{G_2}{G_3}$ and $\beta$. We hace assumed $N=50$ (blue) and $N=60$ (red). To be consistent with the Planck data and BICEP2/Keck Array, $\frac{G_2}{G_3}$ and $\beta$ are constrained inside the coloured areas.}
    \label{Figmodel3}
\end{figure}

Finally, we consider the case where the Hubble rate is given by the following function of the number of e-folds:
    \begin{eqnarray}
       H^2 (N)= G_2 \text{e}^{\beta N}+G_3\ ,
       \label{inflation3_Hn}
    \end{eqnarray}
where the constants $G_2>0$ and $G_3>0$. Following the same procedure as in the previous cases, the equation for $g(N)$ is obtained by inserting (\ref{inflation3_Hn}) in the equation (\ref{gt}), which turns out:
     \begin{eqnarray}
      &&\tilde{C}\text{e}^{3-3N}+\text{e}^{N\beta}G_2\beta(\text{e}^{N\beta}G_2+G_3)\nonumber\\
      &&+4(\text{e}^{N\beta}G_2+G_3)^3(G_3+\text{e}^{N\beta}G_2(-1+3\beta))g(N)\nonumber\\
      &&+4(\text{e}^{N\beta}G_2+G_3)^4g'(N)=0
       \label{eom gn3}
    \end{eqnarray}
Before solving this equation, we can constrain the parameters $G_2$, $G_3$ and $\beta$. Note that now the slow-roll parameters read,
    \begin{eqnarray}
        \label{slow roll 31}
        \epsilon&=&-\frac{\beta G_2\text{e}^{\beta N}( G_2\text{e}^{\beta N}+G_3)}{2( G_2\text{e}^{\beta N}+\frac{6G_3}{6+\beta})^2}\\
        \label{slow roll 32}
        \eta&=&-\frac{\beta ( 2G_2\text{e}^{\beta N}+G_3)}{2( G_2\text{e}^{\beta N}+\frac{6G_3}{6+\beta})}
    \end{eqnarray}
Here, the spectral index $n_s$ and the tensor-to-scalar ratio $r$ (\ref{2.5}) depend on the number of e-folds, the parameter $\beta$ and the ratio $G_2/G_3$, so in order to obtain some information about the value of free parameters by using the  Planck data and BICEP2/Keck Array data as before, we consider two samples, depending on the number of e-folds during inflation. In Fig.~\ref{Figmodel3}, we consider the cases $N=50$ and $N=60$ respectively.
As shown in the figure, parameters are well constrained, particularly the value of the exponent $\beta$. \\

In order to reconstruct the gravitational action analytically, we assume some approximations that can simplify the equations. Otherwise, the mimetic Gauss-Bonnet Lagrangian is not possible to be obtained exactly for this case, as above. Thus, as $\beta\sim 0.03$ seems to be natural and combined with the slow-roll condition $|\dot{H}|\ll H^2$, we obtain $\frac{G_2 \text{e}^{\beta N}}{G_3} \gg 1$, such that the Hubble parameter Eqs. (\ref{inflation3_Hn}) can be approximated as follows 
   \begin{eqnarray}
      && H^2 (N)\simeq G_2 \text{e}^{\beta N}\ .
       \label{inflation31 Hn}
       \eea
And the Gauss-Bonnet invariant is given by:
  \begin{eqnarray}
      {\cal G}(N)\simeq-24(1+\beta)G_2^4 \text{e}^{4N\beta}\ .
      \label{GB3}
    \end{eqnarray}
Then, Eq.~(\ref{eom gn3}) takes the form:
    \begin{eqnarray}
\tilde{C}\text{e}^{3-3N}+\text{e}^{n\beta}G_2\beta\text{e}^{N\beta}G_2
      +4\text{e}^{4N\beta}G_2^4(-1+3\beta)g(N)+4\text{e}^{N\beta}G_2^4g'(N)\simeq0\ .
       \label{eom gn31}
    \end{eqnarray}
By following the same procedure as above, the approximate form for $f({\cal G})$ leads to
    \begin{eqnarray}
      f({\cal G}(N))\sim24G_2^3\beta(1-\beta)\left[\right.
      \frac{\text{e}^{N\beta}}{3G_2^2\beta(\beta-1)}
      +\frac{\tilde{C}\text{e}^{3-3N-N\beta}}{G_2^4(3+\beta)(4+\beta)(3+5\beta)}
      +\frac{c_1\text{e}^{N}}{1-4\beta}
      \left.\right]\ .
       \label{fg3}
    \end{eqnarray}
which in terms of the Gauss-Bonnet invariant leads to: 
      \begin{eqnarray}
      f({\cal G}(N))&=&24G_2^3\beta(1-\beta)\left\{
      \frac{\left[-{\cal G}/24(1+\beta)\right]^{1/4}}{3G_2^3\beta(\beta-1)}
     +\tilde{C}\text{e}^{3}\frac{\left[-{\cal G}/24G_2^4(1+\beta)\right]^{-(3+\beta)/4\beta}}{G_2^4(3+\beta)(4+\beta)(3+5\beta)} \right. \\ \nonumber
     &&  \left.+\frac{c_1\left[-{\cal G}/24G_2^4(1+\beta)\right]^{1/4\beta}}{1-4\beta}
     \right\}\ .
       \label{fg3}
    \end{eqnarray}
And the explicit form for the mimetic Gauss-Bonnet Lagrangian is obtained.

\section{Extensions of mimetic f({\cal G}) gravity}
\label{extensions}
Let us now consider some extensions of mimetic f({\cal G}) gravity and the reconstruction of inflationary models within such extensions. An alternative way of representing mimetic gravity lies on the use of a constraint equation that can be expressed through a Lagrange multiplier in the action \cite{Golovnev:2013jxa,Astashenok:2015haa}:
    \begin{eqnarray}
        S=\int d^4x\sqrt{-g}\left[ \frac{R}{2\kappa^2} + f(\mathcal{G})+\lambda( g^{\mu\nu}\partial_{\mu}\phi\partial_{\nu}\phi+1) \right]\ .
        \label{action_mgb21}
    \end{eqnarray}
The variation with respect to the Lagrange multiplier $\lambda$ recovers (\ref{Constraint}). Then, it is straightforward to show that the field equations for mimetic $f({\cal G})$ gravity (\ref{eom_metric}) are recovered. Nevertheless, here we are interested to analyse some extensions for the mimetic action (\ref{action_mgb21}) by adding a kinetic term and a potential for the scalar field to make it dynamical:
    \begin{eqnarray}
        S=\int d^4x\sqrt{-g}\left[ \frac{R}{2\kappa^2} + f(\mathcal{G})-\epsilon g^{\mu\nu}\partial_{\mu}\phi\partial_{\nu}\phi-V(\phi)+\lambda( g^{\mu\nu}\partial_{\mu}\phi\partial_{\nu}\phi+1) \right],
        \label{action_mgb2}
    \end{eqnarray}
By assuming again the spatially flat FLRW metric (\ref{metric}), the variation of the action (\ref{action_mgb2}) with respect to the metric $g_{\mu\nu}$ gives rise to the following equations of motion:
    \begin{eqnarray}
        \label{eom21}
        3H^2+24H^{3}\frac{df_{{\cal G}}({\cal G})}{dt}+f({\cal G})-f_{{\cal
        G}}({\cal G}){\cal
        G}=\epsilon \dot{\phi}^{2}+V(\phi)-\lambda(\dot{\phi}^{2}+1), \\
        \label{eom22}
        -2\dot{H}-3H^2-8H^2\frac{d^{2}f_{{\cal G}}({\cal
        G})}{dt^{2}}-16H(\dot{H}+H^2)\frac{df_{{\cal G}}({\cal
        G})}{dt}+f_{{\cal G}}({\cal G}){\cal G}-f({\cal
        G})=\epsilon \dot{\phi}^{2}+-V(\phi)-\lambda(\dot{\phi}^{2}-1).
    \end{eqnarray}
While the variation with respect to the Lagrange multiplier $\lambda$ yields the constraint for the scalar field $\phi$
    \begin{eqnarray}
        \label{constraint}
       \dot{\phi}^2-1=0.
       \label{constrain1}
    \end{eqnarray}
By redefining the mimetic scalar field as cosmic time $\phi=t$ and by combining Eqs. (\ref{eom21}) and (\ref{eom22}), the scalar potential and the Lagrange
multiplier can be expressed as functions of cosmological time $t$ in terms of the Hubble parameter:
    \begin{eqnarray}
        \label{V_t}
        V(t)&=&2\dot{H}+3H^2+\epsilon \dot{\phi}^{2}+8H^2\frac{d^{2}f_{{\cal
        G}}({\cal G})}{dt^{2}}+16H(\dot{H}+H^2)\frac{df_{{\cal G}}({\cal
        G})}{dt}-f_{{\cal G}}({\cal G}){\cal G}+f({\cal G})\ ,   \\
        \label{lambda_t}
        \lambda(t)&=&\dot{\phi}^{-2}\left(\frac{\rho}{2}+\dot{H}+4H(2\dot{H}-H^{2})\frac{df_{{\cal
        G}}({\cal G})}{dt}+4H^{2}\frac{d^{2}f_{{\cal G}}({\cal
        G})}{dt^{2}}\right)+\epsilon\ .
    \end{eqnarray}
Hence, the particular action (\ref{action_mgb2}) can be easily reconstructed for any cosmological model once the Hubble parameter is provided, as shown below for the examples studied in the previous section. Nevertheless, it is more convenient to express the potential $V$ and the Lagrange multiplier $\lambda$ in terms of the $e$-folding number $N$ instead of the cosmological time $t$, such that the
Eqs.~(\ref{V_t})-(\ref{lambda_t}) can be written as follows:
    \begin{eqnarray}
        \label{V_N}
        V(N)&=&2H(N)H'(N)+3H(N)^2+\epsilon H(N)^2 \dot{\phi}^{2}
        +8H(N)^2\left[H(N)^2\frac{d^{2}}{dN^2}+H(N)H'(N)\frac{d}{dN}\right]f_{{\cal G}}({\cal G})\nonumber\\
        &&+16H(N)^3\left[H'(N)+H(N)\right]\frac{df_{{\cal G}}({\cal
        G})}{dN}-f_{{\cal G}}({\cal G}){\cal G}+f({\cal G})\ ,   \\
        \label{lambda_N}
        \lambda(N)&=&\frac{\rho}{2}+4H(N)H'(N)+H(N)^3\left[2H'(N)-H(N)\right]\frac{df_{{\cal G}}({\cal
        G})}{dN}\nonumber\\
        &&+4H(N)^2\left[H(N)^2\frac{d^{2}}{dN^2}+H(N)H'(N)\frac{d}{dN}\right]f_{{\cal G}}({\cal G})+\epsilon\ .
    \end{eqnarray}
Alternatively, the function $f({\cal G})$ in the action (\ref{action_mgb2}) can be expressed in terms of an extra auxiliary field by rewriting the action as follows:
 \begin{eqnarray}
        S=\int d^4x\sqrt{-g}\left[ \frac{R}{2\kappa^2} +\varphi{\cal G}-U(\varphi)-V(\phi)+\lambda( g^{\mu\nu}\partial_{\mu}\phi\partial_{\nu}\phi+1) \right],
        \label{action_mgb3}
    \end{eqnarray}
where $\varphi=f'({\cal G})$ and we have omitted the kinetic term for $\phi$ for simplicity. The variation of the action with respect to $g_{\mu\nu}$ gives the following field equations:
    \begin{eqnarray}
        R_{\mu\nu}- \frac{1}{2}R g_{\mu\nu}+U(\varphi)g_{\mu\nu}+2H_{\mu\nu}
        -\left[\lambda( g^{\mu\nu}\partial_{\mu}\phi\partial_{\nu}\phi+1)-V(\phi)\right]g_{\mu\nu}
        +2\lambda\partial_{\mu}\phi\partial_{\nu}\phi =0\ .
        \label{action_bdm}
    \end{eqnarray}
Here the term $H_{\mu\nu}$ is the variation of $\varphi{\cal G}$,
    \begin{eqnarray}
        H_{\mu\nu} &\equiv& \frac{1}{\sqrt{-g}}\frac{\delta(\sqrt{-g}\varphi{\cal G})}{\delta g^{\mu\nu}} \nonumber\\
        &=&2R(g_{\mu\nu}\Box-\nabla_{\mu}\nabla_{\nu})\varphi
        +4R_{\mu}^{\alpha}\nabla_{\alpha}\nabla_{\nu}\varphi
        +4R_{\nu}^{\alpha}\nabla_{\alpha}\nabla_{\mu}\varphi    \nonumber\\
        &&-4R_{\mu\nu}\Box\varphi
        -4g_{\mu\nu}R^{\alpha\beta}\nabla_{\alpha}\nabla_{\beta}\varphi
        +4R_{\alpha\mu\beta\nu}\nabla^{\alpha}\nabla^{\beta}\varphi
    \end{eqnarray}
Whereas the constrain of the mimetic scalar is given by (\ref{constrain1}). The FLRW equations for the action (\ref{action_bdm}) turn out to be
    \begin{eqnarray}
        &&3H(t)^2(1+8H(t)\varphi'(t))+\lambda(t)(1+\phi'(t)^2)-V(\phi(t))-U(\varphi(t))=0
        \label{bdm_eom1}\\
        &&H(t)\left[ 16H(t)^2\varphi'(t)+16\dot{H}(t)\varphi'(t)+H(t)(3+8\varphi''(t)) \right]
        \nonumber\\
        &&+\lambda(t)\left[ 1-\phi'(t)^2 \right] +2\dot{H}(t)-V(\phi(t))-U(\varphi(t))=0\ .
    \end{eqnarray}
It is straightforward to obtain the Hubble parameter from Eq.~(\ref{bdm_eom1}), leading to the following solution:
    \begin{eqnarray}
        H(N)=\sqrt{\pm\frac{\sqrt{9-96 \varphi '(N) (2 \lambda (N)-U(\varphi (N))-V(\phi (N)))}}{48 \varphi '(N)}-\frac{1}{16 \varphi '(N)}}\ .
    \end{eqnarray}
In the following, the full action (\ref{action_mgb2}) is reconstructed for some inflationary models. Since the action (\ref{action_mgb2}) keeps more degrees of freedom, we are assuming a particular ansazt for the function  $f({\cal G})$, the simplest possible choice which is given by:
    \begin{eqnarray}
       f({\cal G})=A {\cal G}^{2}
    \end{eqnarray}
    Then, by the expressions (\ref{V_N}) and (\ref{lambda_N}), the corresponding action is recovered.\\ 

Let us first consider a simple example, a subclass of the first case studied above, where the cosmological evolution is described by the following Hubble parameter:
    \begin{eqnarray}
       H^2 (N)= G_0 N+G_1,
       \label{inflation1_Hn}
    \end{eqnarray}
Substituting Eq. (\ref{inflation1_Hn}) into Eqs. (\ref{V_N}) and (\ref{lambda_N}) and recalling that $\phi=t$, we obtain
the potential for the scalar field in terms of the number of e-folds:
    \begin{eqnarray}
        V(N)&=& -768 A G_0^2 ( G_0 N+ G_1)^2-12 A  G_0^2 (4  G_0 N+ G_0+4 G_1)\nonumber\\
        &&-192 A  G_0 ( G_0 N+ G_1) (2  G_0 N+ G_0+2  G_1) (4  G_0 N+ G_0+4  G_1)\nonumber\\
        &&-288 A ( G_0 N+ G_1)^2 (2  G_0 N+ G_0+2  G_1)^2-24 A ( G_0 N+ G_1) (2  G_0 N+ G_0+2  G_1)\nonumber\\
        &&+3 ( G_0 N+ G_1)+ G_0+\epsilon\ ,
        \label{VN1}
    \end{eqnarray}
while the configuration of the lagrange multiplier yields:
     \begin{eqnarray}
        \lambda(N)&=&\frac{1}{2} G_0 (1 +
    96 A (G_1 + G_0 N) (8 G_1^2 + 2 G_0 G_1 (-9 + 8 N)\nonumber\\
        && +
       G_0^2 (-3 + 2 N (-9 + 4 N)))) + \epsilon\ .
    \end{eqnarray}
The scalar field as a fucntion of $N$ is obtained by using $\frac{dN}{dt}=H(N)$ and $\phi=t$, leading to
    \begin{eqnarray}
       \phi(N)=-\frac{2(\sqrt{G_0+G_1}-\sqrt{G_1+G_0 N})}{G_0}.
       \label{phiN1}
    \end{eqnarray}
Hence, by combining Eqs. (\ref{VN1}) and (\ref{phiN1}), we obtain the potential $V(\phi)$
\begin{eqnarray}
V(\phi)&=&-768 A  G_0^2 \left(\frac{ G_0^2 \phi ^2}{4}+ G_0 \phi  \sqrt{ G_0+ G_1}+ G_0+ G_1\right)^2   \nonumber\\
        &&-12 A  G_0^2 \left( G_0 \left(4 \phi  \sqrt{ G_0+ G_1}+ G_0 \phi ^2+5\right)+4  G_1\right)   \nonumber\\
        &&-24 A  G_0 \left( G_0 \left(4 \phi  \sqrt{ G_0+ G_1}+ G_0 \phi ^2+4\right)+4  G_1\right) \left( G_0 \left(4 \phi  \sqrt{ G_0+ G_1}+ G_0 \phi ^2+5\right)+4  G_1\right) \nonumber\\
        &&\times  \left( G_0 \left(4 \phi  \sqrt{ G_0+ G_1}+ G_0 \phi ^2+6\right)+4  G_1\right)   \nonumber\\
        &&-\frac{9}{2} A \left( G_0 \left(4 \phi  \sqrt{ G_0+ G_1}+ G_0 \phi ^2+4\right)+4  G_1\right)^2\left( G_0 \left(4 \phi  \sqrt{ G_0+ G_1}+ G_0 \phi ^2+6\right)+4  G_1\right)^2 \nonumber\\
        &&-3 A \left( G_0 \left(4 \phi  \sqrt{ G_0+ G_1}+ G_0 \phi ^2+4\right)+4  G_1\right) \left( G_0 \left(4 \phi  \sqrt{ G_0+ G_1}+ G_0 \phi ^2+6\right)+4  G_1\right)  \nonumber\\
        &&+3 \left(\frac{ G_0^2 \phi ^2}{4}+ G_0 \phi  \sqrt{ G_0+ G_1}+ G_0+ G_1\right)+ G_0+\epsilon.
\end{eqnarray}
 \\
Finally, let us consider the example 3 studied above, where recall that the cosmological evolution is given by $H^2 (N)=-G_2 \text{e}^{\beta N}+G_3$. Similarly to the previous case, the scalar potential and the Lagrange multiplier in terms of the number of e-folds $N$ is obtained
 \begin{eqnarray}
        V(N)&=&-96 A (\beta +2) \left(\beta  (12 \beta +11)+6\right)  G_2^4 e^{4 \beta  N})  \nonumber\\
        &&-24 A G_2^3 \left((\beta +2) \beta ^2+8 (\beta  (\beta  (12 \beta +41)+42)+24) G_3\right) e^{3 \beta  N})  \nonumber\\
        &&-12 A  G_2^2 \left(2 (\beta +2)+8 (\beta  (\beta  (14 \beta +59)+84)+72)  G_3^2+\beta ^2 (\beta +4)  G_3\right) e^{2 \beta  N})  \nonumber\\
        &&- G_2 e^{\beta  N} \left(24 A (\beta +4)  G_3 \left(8 (\beta  (\beta +2)+6) G_3^2+1\right)-\beta -3\right))  \nonumber\\
        &&-3  G_3 \left(16 A \left(24  G_3^3+ G_3\right)-1\right)+\epsilon\ ,
\label{VN2}\\
        \lambda(N)&=&-96 A (\beta +2) (\beta  (12 \beta +11)+6)  G_2^4 e^{4 \beta  N})  \nonumber\\
        &&-24 A  G_2^3 \left((\beta +2) \beta ^2+8 (\beta  (\beta  (12 \beta +41)+42)+24) G_3\right) e^{3 \beta  N})  \nonumber\\
        &&-12 A  G_2^2 \left(2 (\beta +2)+8 (\beta  (\beta  (14 \beta +59)+84)+72)  G_3^2+\beta ^2 (\beta +4)  G_3\right) e^{2 \beta  N})  \nonumber\\
        &&- G_2 e^{\beta  N} \left(24 A (\beta +4)  G_3 \left(8 (\beta  (\beta +2)+6) G_3^2+1\right)-\beta -3\right))  \nonumber\\
        &&-3  G_3 \left(16 A \left(24 G_3^3+ G_3\right)-1\right)+\epsilon\ ,
\label{lambdaN2}
    \end{eqnarray}
whereas the scalar field yields:
 \begin{eqnarray}
\phi(N)&=&\frac{2~\text{arctanh}\left(\frac{\sqrt{\text{e}^\beta G_2+G_3}}{\sqrt{G_3}}\right)-2~\text{arctanh}\left(\frac{\sqrt{\text{e}^{\beta N} G_2+G_3}}{\sqrt{G_3}}\right)}{\beta\sqrt{G_3}}\ .
    \end{eqnarray}
And finally the potential $V(\phi)$ can be expressed as a function of $\phi$ as follows:
     \begin{eqnarray}
       V(\phi)&=& -96 A  G_3^4 (W(\phi)-1) \left(\left(12 \beta ^3+35 \beta ^2+28 \beta +12\right) W(\phi)^3  \right.\nonumber\\
&&\left.-\left(12 \beta ^3+47 \beta ^2+56 \beta +36\right) W(\phi)^2+2 \left(\beta ^3+6 \beta ^2+14 \beta +18\right) W(\phi)-12\right) \nonumber\\
&&+12 A \beta ^2  G_3^3 W(\phi)^2 (-\beta +2 (\beta +2) W(\phi)-4) \nonumber\\
&&-24 A  G_3^2 (W(\phi)-1) ((\beta +2) W-2)- G_3 ((\beta +3) W(\phi)-3)+\epsilon
    \end{eqnarray}
where $W(\phi)=\text{sech}\left(\frac{1}{2}\sqrt{G_3}\beta\phi-\text{arctanh}\left(\frac{\sqrt{\text{e}^\beta G_2+G_3}}{\sqrt{G_3}}\right)\right)^2$. Hence, we have shown that any inflationary model can be easily obtained from the mimetic action (\ref{action_mgb2}).

\section{Conclusions}
\label{conclusions}

As shown previously in the literature, Gauss-Bonnet gravity can easily reproduce any model of inflation by assuming the appropriate gravitational action. Here we have extended such analysis by assuming the mimetic condition, which isolates the conformal degree of freedom through an auxiliary scalar field that behaves as a pressureless fluid in homogeneous and isotropic cosmologies. Then, we have assumed the Hilbert-Einstein action with a correction in the form of a function of the Gauss-Bonnet invariant. By some analytical techniques, we have reconstructed the appropriate $f({\cal G})$ actions for  some inflationary models, which reproduce slow-roll inflation. The predictions of such models have been also analysed by obtaining the values for the spectral index and the tensor-to-scalar ratio. As shown by the results, model 1 provides an spectral index a bit larger than the one provided by Planck, while models 2 and 3 fit correctly the observational constraints, despite the analytical form of the gravitational action is not possible to be obtained for the former but it does for the latter. In comparison to standard Gauss-Bonnet gravities, mimetic gravity introduces an additional degree of freedom that behaves as a pressureless fluid, leading to a more complexity of the equations, which avoids to reconstruct the exact gravitational action for some cases. Nevertheless, we have shown that even with the presence of such a fluid, inflation can be realised in mimetic Gauss- Bonnet gravity. \\

We have also explored some analytical extensions of mimetic Gauss-Bonnet gravity by adding a kinetic term and a potential to the auxiliary scalar field. For that aim, we have used the approach of the Lagrange multiplier, equivalent to the mimetic condition. Note that as the auxiliary field becomes dynamical and the $f({\cal G})$ action can be expressed in terms of another scalar field, inflation becomes equivalent to a multifield model, where isocurvature (non-adiabatic) perturbations may not be null \cite{Sasaki:1995aw}, an aspect that required a further and exclusive analysis in a future work. Nevertheless, by analysing the same inflationary models, the corresponding scalar potential has been reconstructed, where we have assumed $f({\cal G})\propto {\cal G}^2$ for simplicity.  \\

Finally, an interesting point to be considered would be the analysis of loop quantum cosmology corrections within mimetic modified gravities. As shown in \cite{Kleidis:2017ftt} for classical extensions of GR, such corrections may discard some inflationary/bouncing models, providing some more accurate gravitational actions for the corresponding cosmological solutions.\\ 

Hence, we have shown that inflation can be realised in Gauss-Bonnet gravity when incorporating the mimetic condition and viable inflationary models that satisfy the last observational constraints can be reconstructed, which opens new ways to explore the early universe in the framework of this type of theories.

\section*{Acknowledgements}
D.S-C.G. is supported by the Juan de la Cierva program (Spain) No.~IJCI-2014-21733 and project FIS2016-76363-P (Spain). Y.Z. would like to thank the support of the scholarship granted by the Chinese Scholarship Council (CSC). This paper is based upon work from CANTATA COST (European Cooperation in Science and Technology) action CA15117,  EU Framework Programme Horizon 2020.


\end{document}